\documentclass[onecolumn,preprint,showpacs,superscriptaddress]{revtex4-1}

\usepackage[utf8]{inputenc}
\usepackage{mathtools}
\usepackage{graphicx}
\usepackage{caption}
\usepackage{subcaption}
\usepackage{xcolor}
\usepackage{soul}
\usepackage{float}
\usepackage{amsmath}
\usepackage{amssymb}
\usepackage{amsfonts}
\usepackage{xcolor}
\usepackage{appendix}
\usepackage{orcidlink}

\begin{document}

\title{Cornell and Coulomb potentials in the double defect spacetime}

\author{L. G. Barbosa \orcidlink{0009-0007-3468-3718}}
\email{leonardo.barbosa@posgrad.ufsc.br}
\affiliation{Departamento de Física, CFM - Universidade Federal de Santa Catarina; C.P. 476, CEP 88.040-900,\\ Florianópolis, SC, Brazil }

\author{L. C. N. Santos \orcidlink{0000-0002-6129-1820}}
\email{luis.santos@ufsc.br}
\affiliation{Departamento de Física, CFM - Universidade Federal de Santa Catarina; C.P. 476, CEP 88.040-900,\\ Florianópolis, SC, Brazil }

\author{J. V. Zamperlini \orcidlink{0009-0002-9702-1555}}
\email{joao.zamperlini@posgrad.ufsc.br}
\affiliation{Departamento de Física, CFM - Universidade Federal de Santa Catarina; C.P. 476, CEP 88.040-900,\\ Florianópolis, SC, Brazil }

\author{F. M da Silva \orcidlink{0000-0003-2568-2901}}
\email{franciele.m.s@ufsc.br}
\affiliation{Departamento de Física, CFM - Universidade Federal de Santa Catarina; C.P. 476, CEP 88.040-900,\\ Florianópolis, SC, Brazil }

\begin{abstract}
We analyze the quantum dynamics of a scalar field in a spacetime incorporating dual topological defects, specifically a cosmic string and a global monopole. Utilizing a generalized metric that encapsulates the combined geometric effects of both defects, we solve the Klein--Gordon equation through separation of variables and examine the role of external potentials, with a focus on the generalized Cornell potential. A comparative analysis against the pure Coulomb potential is conducted to elucidate the modifications induced by the additional linear term. The presence of topological defects deforms the radial components of the wave equation, leading to energy spectrum shifts in bound states and alterations in scattering phase shifts. The results obtained provide a deeper theoretical foundation for understanding the behavior of spin-0 particles in nontrivial spacetime geometries, particularly in the presence of distinct potential interactions.
\end{abstract}

\maketitle

\section{Introduction}

Cosmic strings and global monopoles are topological defects that can arise in grand unified theories and phase transitions in the early universe. They have both theoretical and observational implications in cosmology and astrophysics \cite{string12,string13,string14}. The gravitational effects associated with such defects can be taken into account in general relativity (GR) and a spacetime metric can be proposed in order to describe the geometry around these objects. In this sense, we can study the classical and quantum dynamics around a topological defect and determine if there is a deviation in the trajectory of test particles in relation to flat spacetime. Besides, such defects are studied not only because of the astrophysical interest but also because of potential applications in condensed matter. In this sense, there is an analogy between disinclination in solids and cosmic strings that allows an application of many results obtained in astrophysical systems in the context of such systems \cite{nelson2002defects,kibble2008introduction}. 

It is well known that the geometry associated with cosmic strings and global monopoles has a metric with a high degree of symmetry enabling exact solutions of the equations of motion.
In particular, the spacetime with defects has been studied in the context of quantum dynamics where the Dirac, Schrödinger, and Klein--Gordon equations have been solved in an exact manner using different techniques. For example, 
non-relativistic systems can be studied in defect spacetime by considering only the spatial part of the spacetime metric of these systems in the Schrödinger equation in curved spaces  \cite{wang2015exact,muniz2014landau,ikot2016solutions,ahmed2023effects}. In this way, the Hamiltonian operator associated with the non-relativistic quantum wave equation depends on the spatial coordinates \cite{santos2023non}. This scenario is different from the case of the Dirac and Klein--Gordon wave equations where their generalized formulations for curved spacetime preserve the covariance of geometry. Considering the description of fermions, the Dirac equation can be generalized to the context of an arbitrary geometry. The procedure involves the choice of a tetrad basis, which may lead to differences in the final wave equation obtained. In the context of cosmic strings, we can highlight several works that solve the Dirac equation in this spacetime \cite{inercial10,inercial8,string8,string7,string1,hosseinpour2015scattering,marques2005exact,bakke2018dirac,hosseinpour2017scattering,cunha2020dirac,lima20192d}. In the case of bosons, the procedure for generalizing the wave equation to curved spaces is simpler and involves replacing the usual derivative with the covariant derivative. Thus, the Klein--Gordon equation in the spacetime of a cosmic string and its solutions have been studied in several papers in the literature in recent years \cite{santos2,string10,string9,string6,string5,string4,string2,neto2020scalar,boumali2014klein,ahmed2021effects,santos2018relativistic}. A similar scenario can be seen in the study of global monopole geometry where solutions of the Schrödinger, Dirac, and Klein--Gordon equations have been addressed in the literature \cite{global4,ahmed2024effects,alves2023approximate,alves2024exact,mustafa2023schrodinger,global2,de2006exact,de2022klein,montigny2021exact,bragancca2020relativistic,ahmed2022relativistic,global3,ali2022vacuum,ren1994fermions,bezerra2001physics}.

On the other hand, the combined effect of a global monopole and a cosmic string in the same geometry has started to be explored very recently \cite{jusufi2015scalar}. The idea is to evaluate the effects of different defects on the geometry. As a result of this approach, the resulting wave equation will be influenced by the parameters associated with both defects. Thus, it is interesting to evaluate the energy spectrum of quantum systems obtained from the solution of the wave equations in these geometries and compare them with results obtained in the geometries of the cosmic string and the global monopole.

In the present work, we advance the study of solutions of the Klein--Gordon equation in the double defect spacetime. We consider the Klein--Gordon equation for a curved spacetime and solve it through the method of separation of variables where the angular coordinates are separated from the radial coordinate. Additionally, we consider scalar, nonminimal, and electromagnetic interactions that depend only on the radial coordinate. In particular, we solve exactly the equations for radial and angular coordinates, in terms of known functions. In the case of the radial equation, in addition to the effects of the defect parameters, we take into account the influence of the parameters of the Cornell and Coulomb potentials. Both, Coulomb~\cite{hosseinpour2015scattering,garcia2015relativistic,wang2015exact,ikot2016solutions,neto2020scalar,string2,string4} and Cornell~\cite{hassanabadi2011cornell,hall2015schrodinger,hosseinpour2018klein,hosseini2019klein,mutuk2019cornell,abu2021bound,ikot2022solutions} potentials have been subject of great interest in the study of wave equations in the literature.

The remainder of this paper is organized as follows: Section \ref{sec2} describes the topological defects considered, namely cosmic strings and global monopoles, and introduces the generalized metric for their combined effects on spacetime. In Section \ref{sec3}, we derive the curved space wave equation and present the generalized Cornell potential and effective potential governing particle interactions. Section \ref{sec4} focuses on the solution for the generalized potential, analyzing its effects on bound and scattering states. We discuss specific cases of scalar and vector potentials, exploring the conditions for bound state existence, and provide numerical implementations with plots illustrating the influence of parameters on energy levels and radial wave function solutions. Finally, Section \ref{sec6} presents our conclusions.
In this paper, we employ Planck units ($G=c=\hbar=1$), which renders all
quantities dimensionless.

\section{Topological Defects and Generalized Metric} \label{sec2}

Topological defects like cosmic strings and global monopoles modify spacetime geometry in notable ways. Their combination can introduce new spacetime features, leading to novel effects on the physical systems described by this geometry. In this section, we examine a generalized spacetime metric that accounts for the effects of both defects and their associated deficit angles.

The metric for a spacetime with a cosmic string and a global monopole in spherical coordinates is given by \cite{bezerra2002bremsstrahlung,jusufi2015scalar}
\begin{equation}
    ds^{2} = -dt^{2} + dr^{2} + \beta^{2} r^{2} d\theta^{2} + \alpha^{2} \beta^{2} r^{2} \sin^{2}\theta d\phi^{2}.
\end{equation}
 
The parameters \(\alpha\) and \(\beta\) modify the angular components of the metric and are associated with the cosmic string and the global monopole, respectively. The deficit solid angle due to the monopole is given by \(\beta^{2} = 1 - 8\pi \tilde{\eta}^{2}\), where \(\tilde{\eta}\) is the symmetry-breaking scale. Similarly, the deficit angle caused by the cosmic string is given by \(\alpha^{2} = (1 - 4\tilde{\mu})^{2}\), where \(\tilde{\mu}\) is the linear mass density of the string. Both parameters affect the spacetime geometry, with \(\beta\) influencing the solid angle deficit and \(\alpha\) determining the conical structure. In the absence of topological defects, they approach unity.

\section{Field Equations and Solutions} \label{sec3}

We begin our analysis of the quantum dynamics of scalar bosons around topological defects by examining the generalized Klein-Gordon equation, which includes scalar and vector potentials as well as nonminimal couplings, written as:
\begin{equation}
    -\frac{1}{\sqrt{-g}}D_{\mu}^{\left(+\right)}\left(g^{\mu\nu}\sqrt{-g}D_{\nu}^{\left(-\right)}\psi\right)+\left(M+V_{s}\right)^{2}\psi=0,
    \label{3.1}
\end{equation}
where $\psi=\psi(t,r,\theta,\phi)$ represents the bosonic field, $g=\det(g_{\mu\nu})$ and the covariant derivative terms in this expression, \(D_{\mu}^{\left(\pm\right)}\), are defined as:
\begin{equation}    D_{\mu}^{\left(\pm\right)}=\partial_{\mu}\pm X_{\mu}-ieA_{\mu},
\label{3.2}
\end{equation}
where the scalar \(V_s\) and vector potentials \(A_{\mu}\) and \(X_{\mu}\) depend only on the radial coordinate, and \(e\) represents the associated electric charge. The potential \(A_{\mu}\) describes the electromagnetic interaction, and \(X_{\mu}\), referred to as the nonminimal vector potential, consists of both a temporal and a radial component, given respectively by:
\begin{gather}
    A_{\mu}=\left(A_{t}\left(r\right),0,0,0\right), \label{3.3} \\
    X_{\mu}=\left(X_{t}\left(r\right),X_{r}\left(r\right),0,0\right). \label{3.4}
\end{gather}   

At this juncture, we aim to find solutions to the Klein--Gordon equation with the assumption that the wave function can be expressed as:
\begin{equation}
\psi\left(t,r,\theta,\phi\right)=\frac{u\left(r\right)}{r}f\left(\theta\right)e^{im\phi}e^{-i\varepsilon t},
\label{3.5} 
\end{equation} 
where $\varepsilon$ is the energy of the particle and $m \in \mathbb{Z}^*$. Substituting Equation~(\ref{3.5}) into expression (\ref{3.1}) reveals that the angular function \(f(\theta)\) satisfies a differential equation incorporating the deficit angle parameter \(\alpha\) related to the cosmic string defect, given by:
\begin{equation}
  \left[\frac{1}{\sin \theta}\frac{d}{d\theta}\left(\sin \theta\frac{d}{d\theta}\right)+\left(\lambda_{\alpha}-\frac{m^{2}}{\alpha^{2}\sin^2 \theta}\right)\right]f\left(\theta\right)=0.
  \label{3.6}
\end{equation}
The quantity \(\lambda_{\alpha}\) corresponds to the eigenvalues of the angular momentum operator in the presence of the defect and is given by \( \lambda_{\alpha} = l_{\alpha} (l_{\alpha} + 1) \), where \( l_{\alpha} = n + |m_{\alpha}| = l + |m|(1/\alpha - 1) \), with \( l = n + |m| \) and \( m_{\alpha} = m/\alpha \). Here, \( n \in \mathbb{N} \) and \( m_{\alpha} \) satisfies the condition \( -l_{\alpha} \leq m_{\alpha} \leq l_{\alpha} \). The parameters \( l \) and \( m \) denote the orbital angular momentum and magnetic quantum numbers in the absence of the defect.

For the radial part \( u(r) \), we derive a differential equation that describes the wave function's behavior in the radial direction, incorporating contributions from both topological defect parameters. The resulting radial equation is expressed as
\begin{equation}
    \frac{d^{2}u\left(r\right)}{dr^{2}}+\left(K^{2}-V_{\mathrm{eff}}^2-\frac{l_{\alpha}\left(l_{\alpha}+1\right)}{\beta^{2}r^{2}}\right)u\left(r\right)=0 ,
    \label{3.8}
\end{equation}
in which the term \(K^{2}=\varepsilon^{2}-M^{2}\) and \(V_{\mathrm{eff}}\) is the effective potential given by 
\begin{equation}\label{3.9}
    V_{\mathrm{eff}}^2 =V_{s}^{2} - e^{2}A_{t}^{2} + 2\left(MV_{s} - e\varepsilon A_{t}\right) + \frac{\partial X_{r}}{\partial r} + \frac{2}{r}X_{r} + X_{r}^{2} - X_{t}^{2}.
\end{equation}

It is important to notice that the effective potential \(V_{\mathrm{eff}}\) incorporates the scalar and vector potentials, along with nonminimal interaction terms. This structure enables the analysis of scalar boson dynamics in the context of cosmic strings and global monopoles in a unified form.

\section{Solutions for Coulomb and Cornell potentials} \label{sec4}

In this work, the vector potential $A_{\mu}$ and the scalar potential $V_s$ will be given by the Coulomb potential
\begin{equation}
  A_{t}=\frac{\gamma_{t}}{r},\quad V_{s}=\frac{\gamma_{s}}{r},
\end{equation}
and the nonminimal vector potential $X_{\mu}$ will be given by the Cornell potential, which is a combination of a Coulomb-like function and a linear potential, as follows:
\begin{equation}
    X_{t}=\frac{\delta_{t}}{r}+\Delta_{t}r,\quad X_{r}=\frac{\delta_{r}}{r}+\Delta_{r}r.
\end{equation}
Here, \(\gamma_t\), \(\gamma_s\), \(\delta_t\), and \(\delta_r\) are constants associated with the inverse linear potentials of the system, while \(\Delta_t\) and \(\Delta_r\) are constants representing the terms corresponding to the linear potentials. In this case, the effective potential can be written as follows:
\begin{align}\label{eqveff}
V_{\text{eff}}^{2} = & \, \Delta_{r}\left(3+2\delta_{r}\right) - 2\Delta_{t}\delta_{t} 
+ \frac{2\left(M\gamma_{s} - e\varepsilon\gamma_{t}\right)}{r} \notag \\
& + \frac{\gamma_{s}^{2} - e^{2}\gamma_{t}^{2} + \delta_{r}\left(\delta_{r}+1\right) - \delta_{t}^{2}}{r^{2}} 
+ \left(\Delta_{r}^{2} - \Delta_{t}^{2}\right)r^{2} .
\end{align}

Considering that \(\delta_{t} \gg \Delta_{t}\) and \(\delta_{r} \gg \Delta_{r}\), we can approximate \(\Delta_{t}^{2} \approx 0\) and \(\Delta_{r}^{2} \approx 0\), simplifying the analysis of the effective potential. Defining the parameters:
\begin{align}
    \mathcal{K}^{2}=K^{2}-\Delta_{r}\left(3+2\delta_{r}\right)+2\Delta_{t}\delta_{t},
\end{align}
\begin{align}
     \alpha_{1} = 2\left(M\gamma_{s} - e\varepsilon\gamma_{t}\right),
\end{align}
\begin{align}
     \alpha_{2} = \delta_{r}\left(\delta_{r}+1\right) + \gamma_{s}^{2} - e^{2}\gamma_{t}^{2} - \delta_{t}^{2},
\end{align}
the radial equation takes the form:
\begin{equation}
  \frac{d^{2}u\left(r\right)}{dr^{2}}+\left(\mathcal{K}^{2}-\frac{\alpha_{1}}{r}-\frac{\alpha_{2}+l_{\alpha}\left(l_{\alpha}+1\right)/\beta^{2}}{r^{2}}\right)u\left(r\right)=0.
    \label{sch}
\end{equation}

This equation can be seen as a Schrödinger-type differential equation. For this effective potential to have a well-defined structure, the condition $\alpha_1 < 0$ must be considered.

In addition, bound states satisfy the relation 
\begin{equation}
    -\left|\tau\right|<\varepsilon<\left|\tau\right|,
\end{equation}
with
\begin{equation}
    \tau^{2}=M^{2}+\Delta_{r}\left(3+2\delta_{r}\right)-2\Delta_{t}\delta_{t}.
\end{equation}

These conditions thus define the allowed energy values for the bound states. By applying the variable transformation \(z = -2i\mathcal{K}r\), Equation~(\ref{sch}) simplifies to:  
\begin{equation}
    \frac{d^{2}u}{dz^{2}}+\left(-\frac{1}{4}-i\frac{\alpha_{1}}{2\mathcal{K}z}-\frac{\alpha_{2}+l_{\alpha}\left(l_{\alpha}+1\right)/\beta^{2}}{z^{2}}\right)u=0.
\end{equation}

To further simplify Equation~(\ref{sch}), we introduce the following parameters:
\begin{equation}
  \eta=\frac{\alpha_{1}}{2\mathcal{K}},  \qquad \gamma_{l}^{2} = \frac{1}{4} + \frac{l_{\alpha}\left(l_{\alpha}+1\right)}{\beta^{2}} + \alpha_{2}.
\end{equation}

With these definitions, the radial equation can be rewritten in the form of Whittaker’s equation:
\begin{equation}
    \frac{d^{2}u\left(z\right)}{dz^{2}} + \left(-\frac{1}{4} - \frac{i\eta}{z} + \frac{\frac{1}{4} - \gamma_{l}^{2}}{z^{2}}\right)u\left(z\right) = 0.
\end{equation}
The general solution to this equation is given by a linear combination of the Whittaker functions:  
\begin{equation}
    u\left(z\right) = A M_{-i\eta,\gamma_{l}}\left(z\right) + B W_{-i\eta,\gamma_{l}}\left(z\right),
\end{equation}
where \( A \) and \( B \) are constants. The Whittaker functions \( M_{-i\eta,\gamma_{l}}\left(z\right) \) and \( W_{-i\eta,\gamma_{l}}\left(z\right) \) are related to Kummer's confluent hypergeometric functions \( M(a,b;z) \) and \( U(a,b;z) \) through  
\begin{align}
    M_{-i\eta,\gamma_{l}}\left(z\right) &= e^{-\frac{1}{2}z}z^{\frac{1}{2}+\gamma_{l}} M\left(\frac{1}{2}+\gamma_{l}+i\eta,1+2\gamma_{l},z\right), \\
    W_{-i\eta,\gamma_{l}}\left(z\right) &= e^{-\frac{1}{2}z}z^{\frac{1}{2}+\gamma_{l}} U\left(\frac{1}{2}+\gamma_{l}+i\eta,1+2\gamma_{l},z\right).
\end{align}
To obtain a well-defined solution at the origin, we impose \( B = 0 \), reducing the expression to  
\begin{equation}
    u\left(z\right) = A e^{-\frac{z}{2}} z^{\frac{1}{2}+\gamma_{l}} M\left(\frac{1}{2}+\gamma_{l}+i\eta, 1+2\gamma_{l}, z\right).
    \label{hiper}
\end{equation}

For large \(|z|\), the asymptotic form of the confluent hypergeometric function \(M(a,b;z)\) is given by  
\begin{equation}
    M\left(a,b;z\right) \simeq  \frac{\Gamma(b)}{\Gamma(b-a)} e^{-\frac{i}{2}\pi a} |z|^{-a} + \frac{\Gamma(b)}{\Gamma(a)} e^{-i\left(|z| + \frac{\pi}{2}(a-b)\right)} |z|^{a-b}.
\end{equation}

This asymptotic expansion provides insight into the wave function's behavior at large distances, which is particularly relevant in the analysis of scattering processes.

\subsection{Scattering States}

For the study of scattering states, we consider the asymptotic limit where \(|z| \gg 1\) and \(\mathcal{K} \in \mathbb{R}\). In this regime, the radial solution takes the form  
\begin{equation}
  u\left(r\right)\simeq\sin\left(\mathcal{K}r-\frac{l\pi}{2}+\delta_{l}\right),
\end{equation}
where the phase shift \(\delta_{l}\) is given by  
\begin{equation}
   \delta_{l}=\frac{\pi}{2}\left(l+\frac{1}{2}-\gamma_{l}\right)+\mathrm{arg}\Gamma\left(\frac{1}{2}+\gamma_{l}+i\eta\right).
\end{equation} 

For spherically symmetric potentials, the scattering amplitude can be expressed as a series of partial waves:  
\begin{equation}
    f\left(\theta\right) = \sum_{l=0}^{\infty} \left(2l+1\right) f_{l} P_{l}\left(\cos \theta \right),
\end{equation}
where \(P_{l}\left(\cos \theta\right)\) are the Legendre polynomials, and \(f_{l}\) represents the contribution of each partial wave.  

The scattering matrix \(S_{l}\) for a given angular momentum quantum number \(l\) is written as  
\begin{equation}
    S_{l} = e^{2i\delta_{l}} = e^{i\pi\left(l+\frac{1}{2}-\gamma_{l}\right)} \frac{\Gamma\left(\frac{1}{2} + \gamma_{l} + i\eta\right)}{\Gamma\left(\frac{1}{2} + \gamma_{l} - i\eta\right)}.
\end{equation}  

The expression for \(S_{l}\) incorporates the phase shifts associated with the Coulomb and Cornell potentials and the topological defects, offering relevant information about the scattering process.

\subsection{Bound States}

The S-matrix exhibits singular behavior under the condition:  
\begin{equation}
    \frac{1}{2} + \gamma_{l} + i\eta = -N,
\end{equation}
where \(N = 0, 1, 2, \dots\). Introducing the parameter:  
\begin{equation}
    \mu = N + \frac{1}{2} + \gamma_{l},
\end{equation}
and substituting the expression for \(\eta\), we arrive at the following quadratic equation for \(\varepsilon\):  
\begin{equation}
   \left(e^{2}\gamma_{t}^{2}+\mu^{2}\right)\varepsilon^{2}-2M\gamma_{s}e\gamma_{t}\varepsilon+M^{2}\gamma_{s}^{2}-\tau^{2}\mu^{2}=0,
\end{equation}
which can be solved to determine the energy spectrum of the bound states:  
\begin{equation}
    \varepsilon_{\pm}=\frac{M}{\left(1+\frac{e^{2}\gamma_{t}^{2}}{\mu^{2}}\right)}\left(\frac{\gamma_{s}}{\mu}\frac{e\gamma_{t}}{\mu}\pm\sqrt{\left(1+\frac{e^{2}\gamma_{t}^{2}}{\mu^{2}}\right)\frac{\tau^{2}}{M^{2}}-\frac{\gamma_{s}^{2}}{\mu^{2}}}\right).
    \label{energia}
\end{equation}

{

To visualize the impact of the potential linear terms on energy levels, we refer to \autoref{fig:oscillator-energy-Deltar-varyDeltat}. As $\Delta_r$ increases, the energy becomes more negative, suggesting that this term enhances bonding in curved spacetime. Conversely, increasing $\Delta_t$ for a fixed $\Delta_r$ raises such energy, thereby weakening the interaction.
%%%%%%%%%%%%%%%%%%%%%%%%%%%%%%%%%%%%%%%%%%%%%%%%%%%%%%%%%%%%%%%%%%%%%%%%%%%%%%%%%%%%%%%%%%%%%%%%%%%%%
\begin{figure}[H]
    \centering
    \includegraphics[scale=0.5]{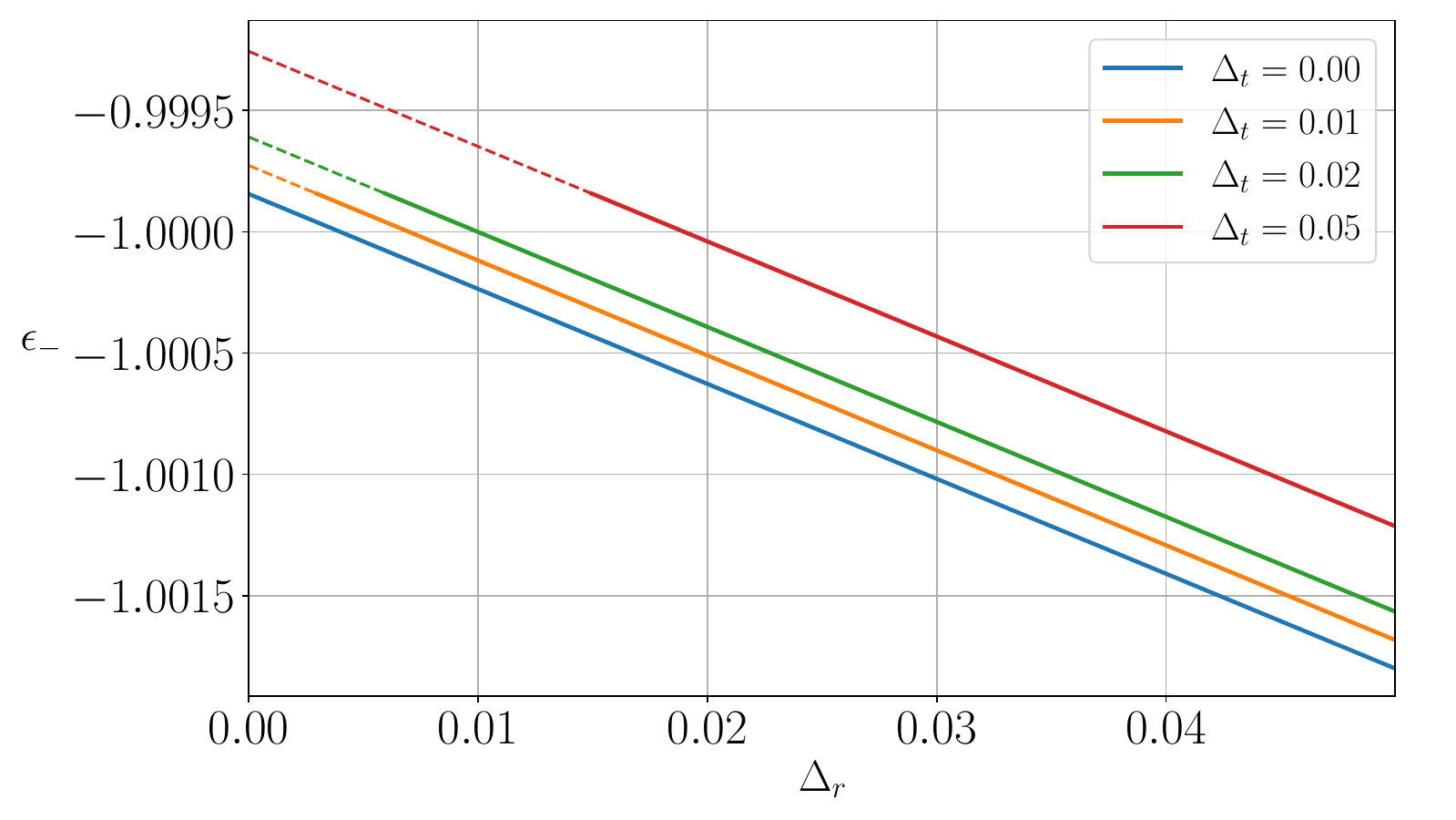}
    \caption{Plot of the negative energy values against the linear potential parameter $\Delta_r$ for four different values of $\Delta_t$, with parameters $\delta_r = 0.5$, $\delta_t = 0.6$, $\gamma_s = 0.7$, $e\gamma_t = -0.8$, $\alpha = 0.9$, $m = l = 1$, $\beta=0.3$, and $M=1$. Solid lines satisfy the bound state condition, while dashed line does not.}
    \label{fig:oscillator-energy-Deltar-varyDeltat}
\end{figure}
%%%%%%%%%%%%%%%%%%%%%%%%%%%%%%%%%%%%%%%%%%%%%%%%%%%%%%%%%%%%%%%%%%%%%%%%%%%%%%%%%%%%%%%%%%%%%%%%%%%%%
We can examine how the linear potential parameters vary with the quantum numbers $N$ and $l$ by referring to \autoref{fig:oscillator-3dplot-Nl}, specifically for the case where $\Delta_r=\Delta_t=0.01$. In this comparison with the scenario of pure Coulomb potential features \cite{barbosa2024}, we find that for $N=0$, there is a significantly greater energy decrease compared to other quantum number configurations, while higher $l$ levels exhibit a smaller decrease.
%%%%%%%%%%%%%%%%%%%%%%%%%%%%%%%%%%%%%%%%%%%%%%%%%%%%%%%%%%%%%%%%%%%%%%%%%%%%%%%%%%%%%%%%%%%%%%%%%%%%%
\begin{figure}[H]
    \centering
    \includegraphics[scale=0.5]{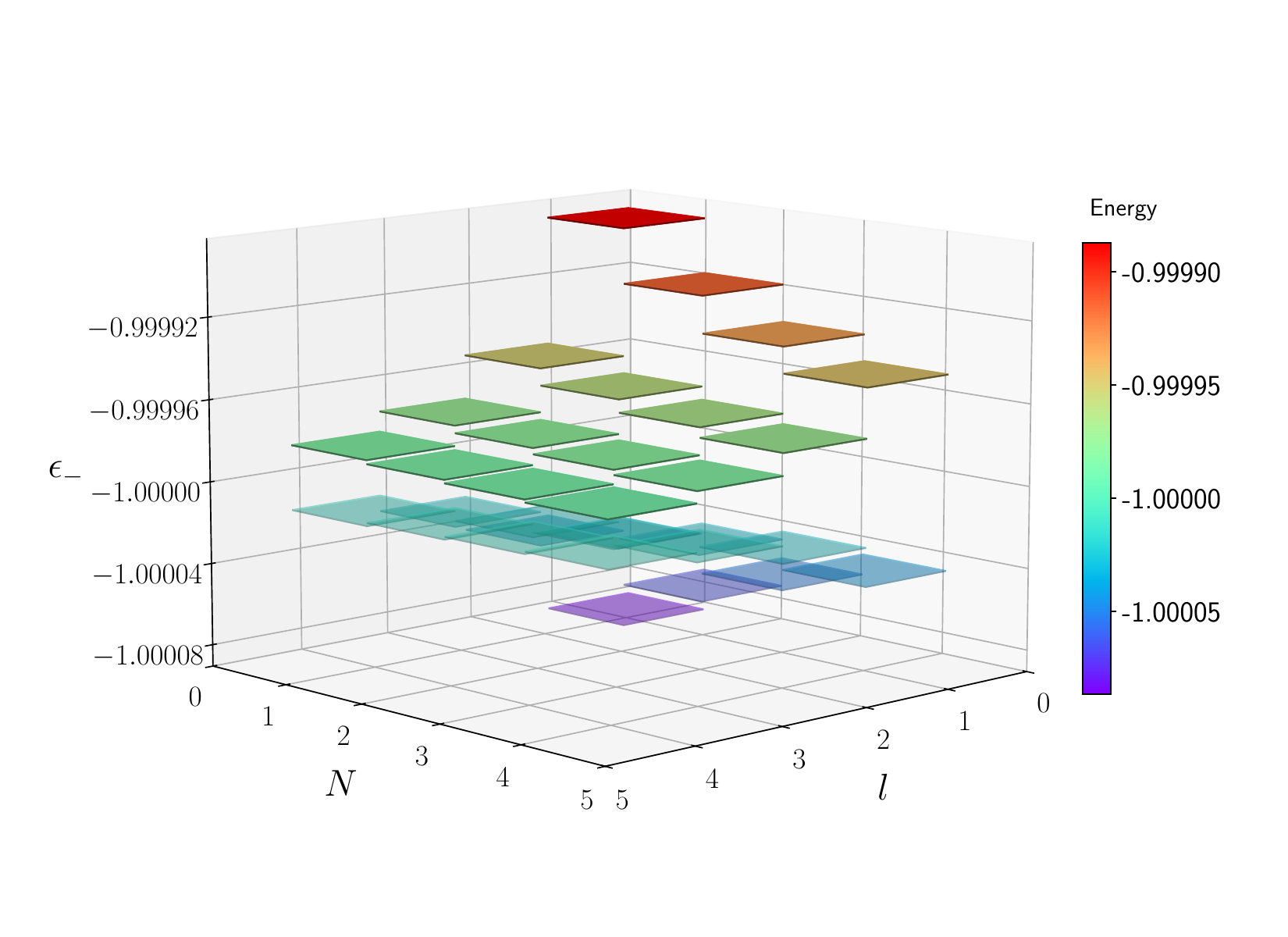}
    \caption{Plot of the negative energy spectrum dependent on the quantum numbers $N$ and $l$, which can take the values \{1,2,3,4\}, for the case of pure Coulomb potential (solid tiles) and Cornell potential with $\Delta_r=\Delta_t=0.01$ (semi-transparent tiles), with parameters $\delta_r = 0.5$, $\delta_t = 0.6$, $\gamma_s = 0.7$, $e\gamma_t = -0.8$, $\alpha = 0.9$, $m = 1$, $\beta=0.3$, and $M=1$.}
    \label{fig:oscillator-3dplot-Nl}
\end{figure}
%%%%%%%%%%%%%%%%%%%%%%%%%%%%%%%%%%%%%%%%%%%%%%%%%%%%%%%%%%%%%%%%%%%%%%%%%%%%%%%%%%%%%%%%%%%%%%%%%%%%%
In the context of comparing with prior studies, it is pertinent to examine how the energy shifts with variations in the cosmic string parameter $\alpha$, particularly across various linear term coefficients configurations. According to \autoref{fig:oscillator-energyvsalpha}, the energy decreases within the regimes under consideration. An intriguing result, however, is the mitigation of a substantial change with $\alpha$ when $\Delta_t=2\Delta_r=0.02$, resulting in the concentration of energy around $-M$.
%%%%%%%%%%%%%%%%%%%%%%%%%%%%%%%%%%%%%%%%%%%%%%%%%%%%%%%%%%%%%%%%%%%%%%%%%%%%%%%%%%%%%%%%%%%%%%%%%%%%%
\begin{figure}[H]
    \centering
    \includegraphics[scale=0.5]{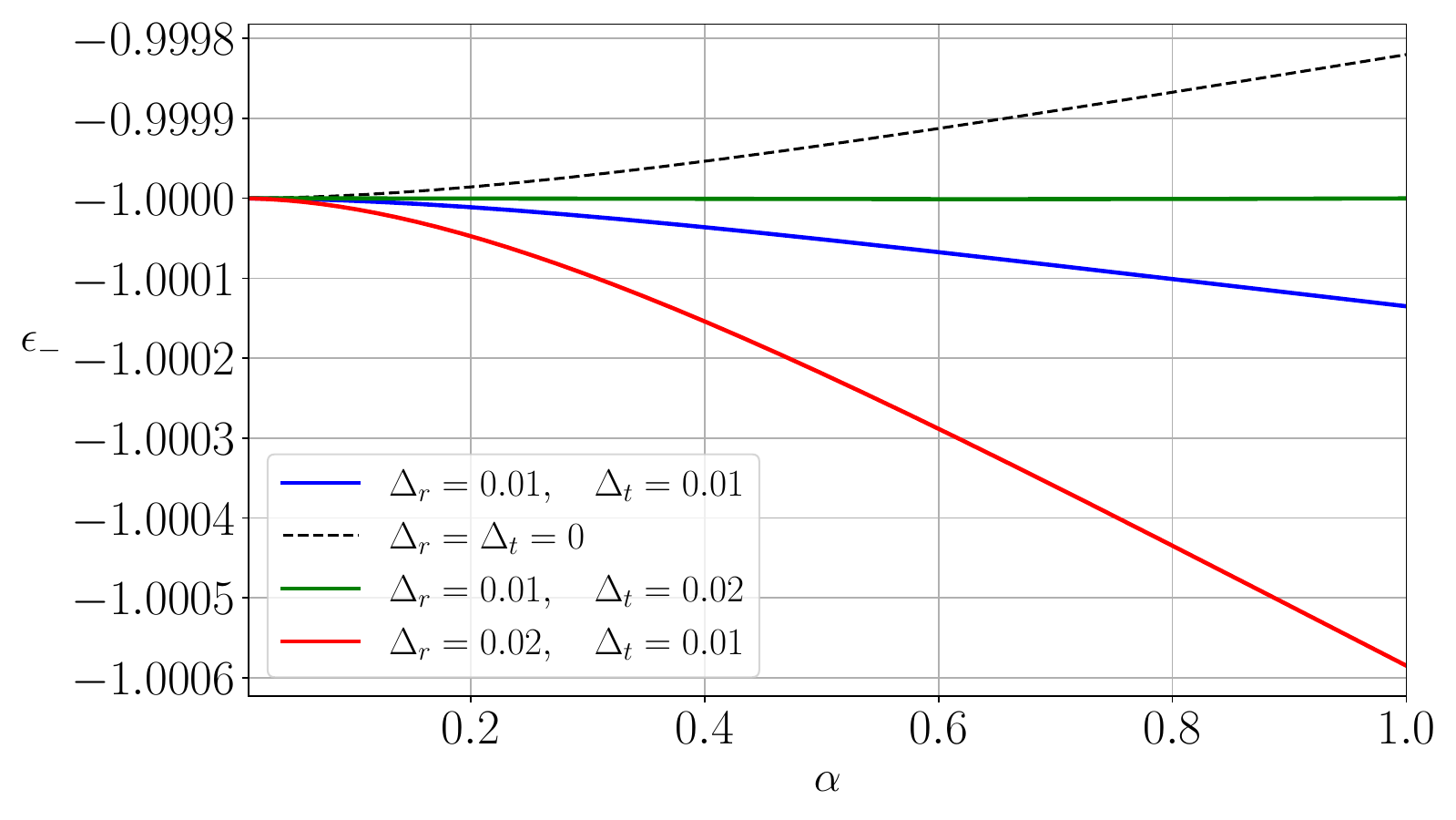}
    \caption{Plots of the $N=0$ energy level against angular deficit, presenting the case of the pure Coulomb interaction given by blacked dashed lines, while three different configurations of linear terms from the Cornell potential are shown using colored solid lines, with parameters $\delta_r = 0.5$, $\delta_t = 0.6$, $\gamma_s = 0.7$, $e\gamma_t = -0.8$, $\alpha = 0.9$, $m = 1$, $\beta=0.3$, and $M=1$.}
    \label{fig:oscillator-energyvsalpha}
\end{figure} 
%%%%%%%%%%%%%%%%%%%%%%%%%%%%%%%%%%%%%%%%%%%%%%%%%%%%%%%%%%%%%%%%%%%%%%%%%%%%%%%%%%%%%%%%%%%%%%%%%%%%%
In addition to the energy values analyzed, it is important to explore how introducing the Cornell potential linear term modifies the radial wave functions at the initial energy levels $N$. According to \autoref{fig:oscillator-radialwavef-energylevels}, there is a significant alteration in the spread of the radial wave functions. Specifically, with $\Delta_r=\Delta_t=0.01$, the established result from \cite{barbosa2024} shows a strong concentration near the origin, at an effective distance more than 10 times smaller. This results in a reduction of the average distance in the new probability distribution, as illustrated in \autoref{fig:oscillator-radialrprob-energylevels}.

Also, we can compare how varying the values of $\Delta_r$ and $\Delta_t$ affect the wave function radial distribution. By considering three different scenarios, we can perceive from Figures \ref{fig:oscillator-radialwavef-DELTAVARY} and \ref{fig:oscillator-radialprob-DELTAVARY} that when $\Delta_r=2\Delta_t=0.02$ (the strengthened bonding scenario) the wave function becomes less spread and more localized towards the origin of the coordinate center. While, when setting the opposite: $\Delta_t=2\Delta_r=0.02$, we can see that the concentration is attenuated by the larger value of $\Delta_t$, illustrating well the role these parameters play in the quantum dynamics of the scalar particle.
%%%%%%%%%%%%%%%%%%%%%%%%%%%%%%%%%%%%%%%%%%%%%%%%%%%%%%%%%%%%%%%%%%%%%%%%%%%%%%%%%%%%%%%%%%%%%%%%%%%%%
\begin{figure}[H]
    \centering
    \includegraphics[scale=0.5]{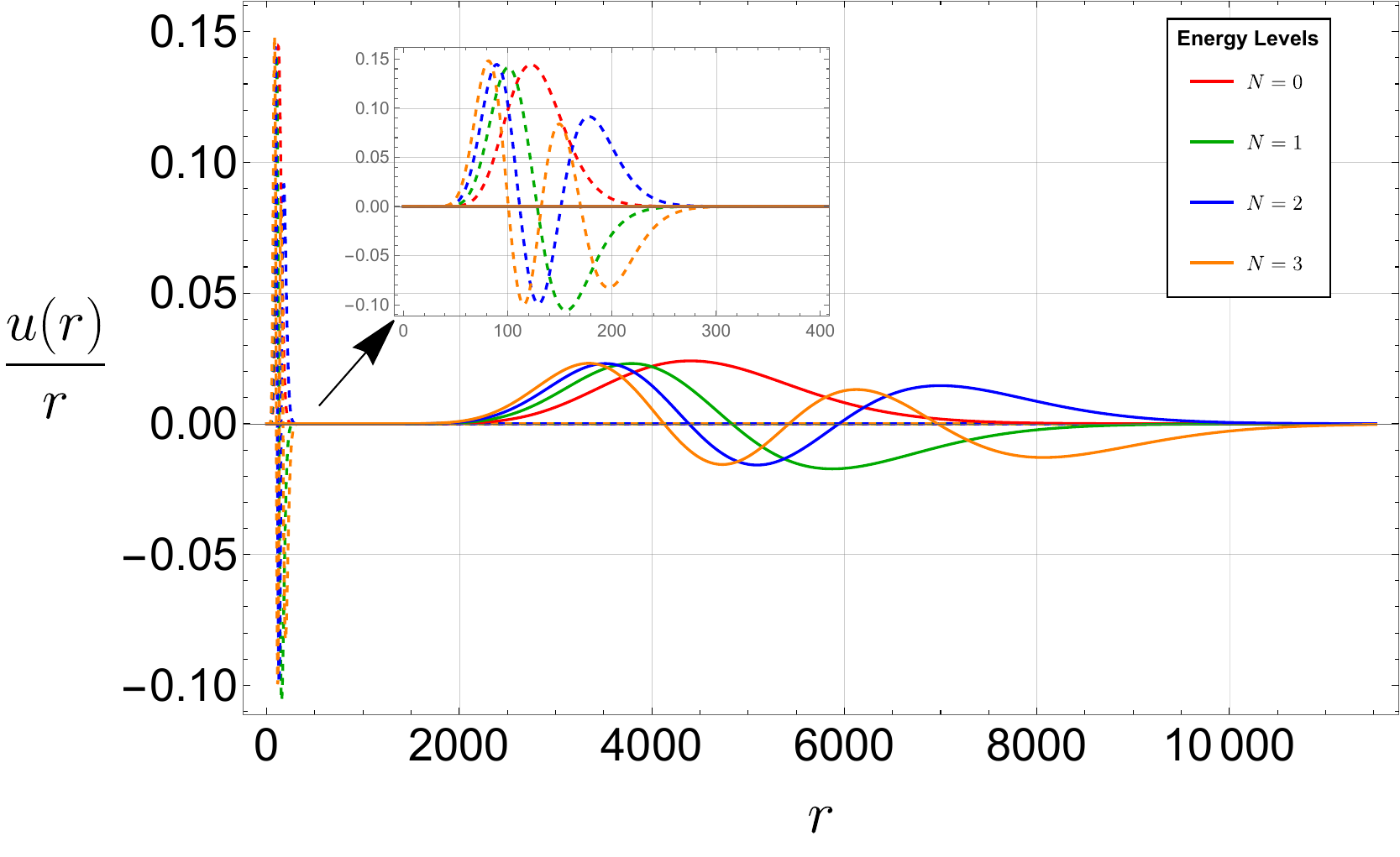}
    \caption{Plot of the radial wave functions for four different values of $N$, for both the case of pure Coulomb interaction (solid lines) and Cornell potential with $\Delta_r=\Delta_t=0.01$ (dashed lines) with parameters $\delta_r = 0.5$, $\delta_t = 0.6$, $\gamma_s = 0.7$, $e\gamma_t = -0.8$, $\alpha = 0.1$, $m = 1$, $l=1$, $\beta=0.5$, and $M=1$.}
    \label{fig:oscillator-radialwavef-energylevels}
\end{figure}
%%%%%%%%%%%%%%%%%%%%%%%%%%%%%%%%%%%%%%%%%%%%%%%%%%%%%%%%%%%%%%%%%%%%%%%%%%%%%%%%%%%%%%%%%%%%%%%%%%%%%
%%%%%%%%%%%%%%%%%%%%%%%%%%%%%%%%%%%%%%%%%%%%%%%%%%%%%%%%%%%%%%%%%%%%%%%%%%%%%%%%%%%%%%%%%%%%%%%%%%%%%
\begin{figure}[H]
    \centering
    \includegraphics[scale=0.5]{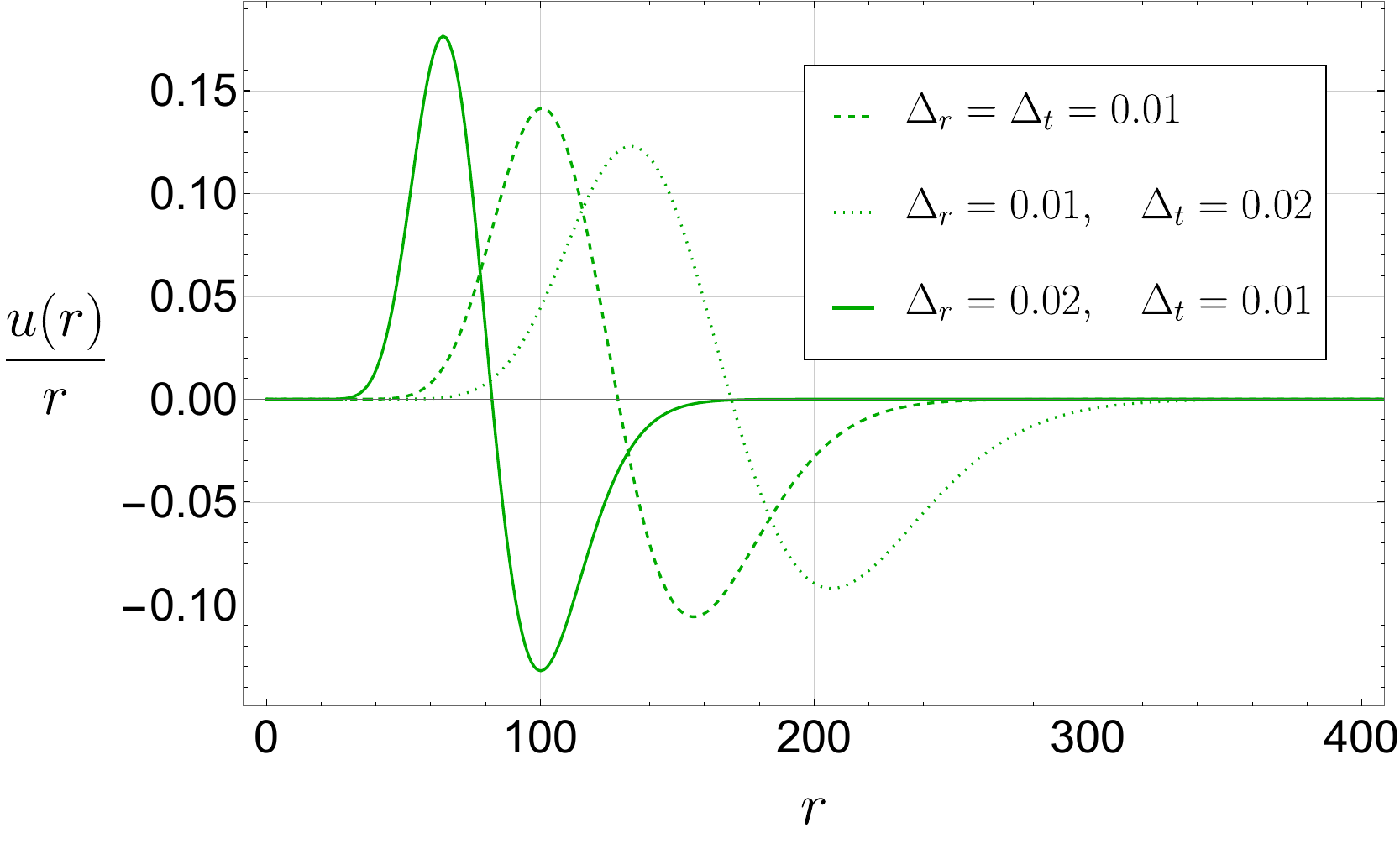}
    \caption{Plot of the radial wave functions for $N=1$, for both three different configurations of linear potential terms: $\Delta_r>\Delta_t$ (solid lines), $\Delta_r=\Delta_t$ (dashed lines) and $\Delta_r<\Delta_t$ (dotted lines) with parameters $\delta_r = 0.5$, $\delta_t = 0.6$, $\gamma_s = 0.7$, $e\gamma_t = -0.8$, $\alpha = 0.1$, $m = 1$, $l=1$, $\beta=0.5$, and $M=1$.}
    \label{fig:oscillator-radialwavef-DELTAVARY}
\end{figure}
%%%%%%%%%%%%%%%%%%%%%%%%%%%%%%%%%%%%%%%%%%%%%%%%%%%%%%%%%%%%%%%%%%%%%%%%%%%%%%%%%%%%%%%%%%%%%%%%%%%%%
%%%%%%%%%%%%%%%%%%%%%%%%%%%%%%%%%%%%%%%%%%%%%%%%%%%%%%%%%%%%%%%%%%%%%%%%%%%%%%%%%%%%%%%%%%%%%%%%%%%%%
\begin{figure}[H]
    \centering
    \includegraphics[scale=0.5]{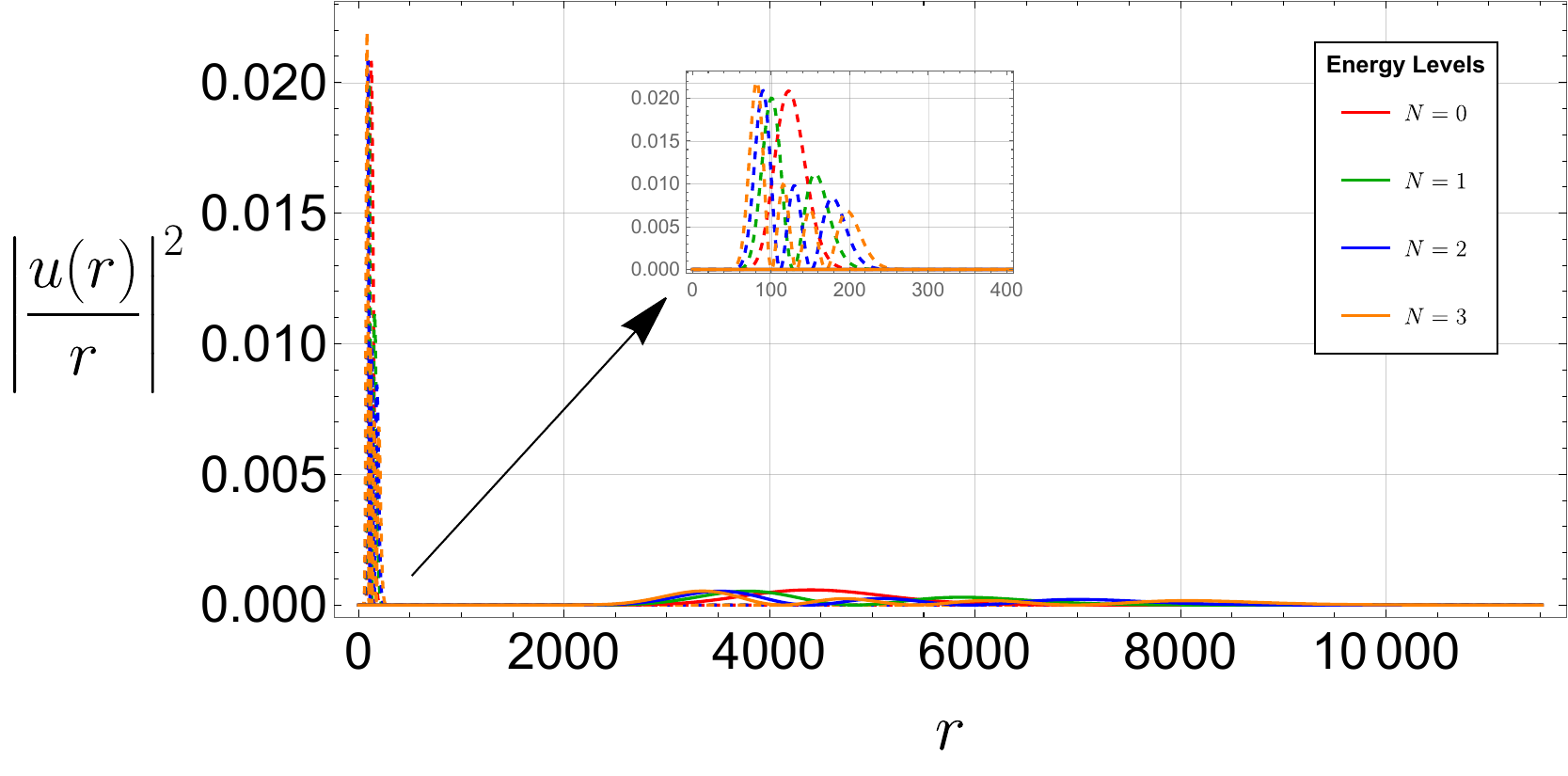}
    \caption{Plot of the radial probability density for $N=1$, for both the case of pure Coulomb potential (solid lines) and Cornell potential with $\Delta_r=\Delta_t=0.01$ (dashed lines) with parameters $\delta_r = 0.5$, $\delta_t = 0.6$, $\gamma_s = 0.7$, $e\gamma_t = -0.8$, $\alpha = 0.1$, $m = 1$, $l=1$, $\beta=0.5$, and $M=1$.}
    \label{fig:oscillator-radialrprob-energylevels}
\end{figure}
%%%%%%%%%%%%%%%%%%%%%%%%%%%%%%%%%%%%%%%%%%%%%%%%%%%%%%%%%%%%%%%%%%%%%%%%%%%%%%%%%%%%%%%%%%%%%%%%%%%%%
%%%%%%%%%%%%%%%%%%%%%%%%%%%%%%%%%%%%%%%%%%%%%%%%%%%%%%%%%%%%%%%%%%%%%%%%%%%%%%%%%%%%%%%%%%%%%%%%%%%%%
\begin{figure}[H]
    \centering
    \includegraphics[scale=0.5]{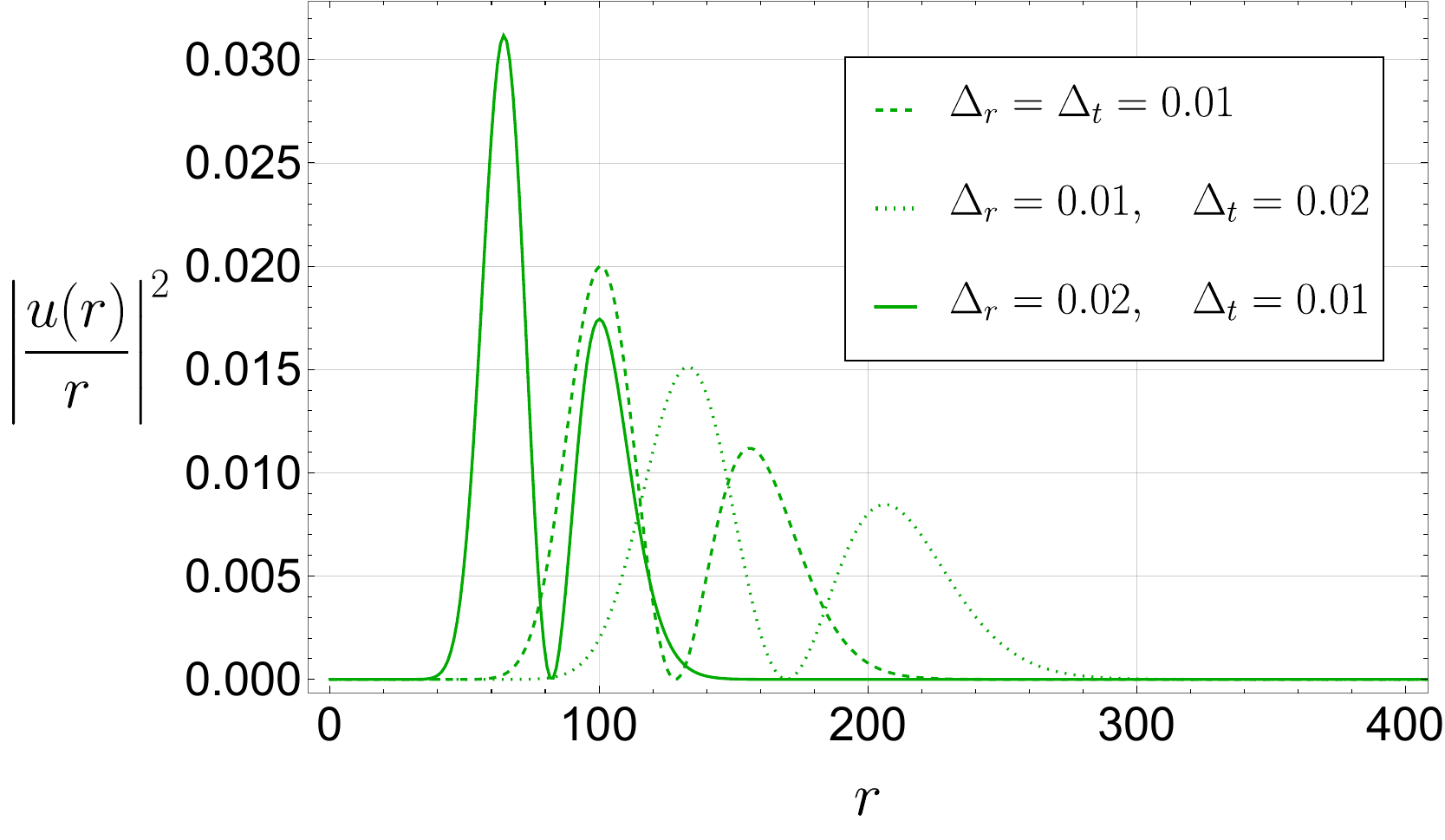}
    \caption{Plot of the radial probability density for $N=1$, for both three different configurations of linear potential terms: $\Delta_r>\Delta_t$ (solid lines), $\Delta_r=\Delta_t$ (dashed lines) and $\Delta_r<\Delta_t$ (dotted lines) with parameters $\delta_r = 0.5$, $\delta_t = 0.6$, $\gamma_s = 0.7$, $e\gamma_t = -0.8$, $\alpha = 0.1$, $m = 1$, $l=1$, $\beta=0.5$, and $M=1$.}
    \label{fig:oscillator-radialprob-DELTAVARY}
\end{figure}
}
%%%%%%%%%%%%%%%%%%%%%%%%%%%%%%%%%%%%%%%%%%%%%%%%%%%%%%%%%%%%%%%%%%%%%%%%%%%%%%%%%%%%%%%%%%%%%%%%%%%%%

\subsection{Particular Cases} \label{sec5}

As a consistency test for our results, we can examine particular cases of the obtained energy spectrum. The first case of interest $\Delta_t = \Delta_r = 0$ corresponds to the Cornell potential with the linear term absent, \textit{i.e.}, a Coulomb-like potential. In this case, the energy spectrum reduces to 
\begin{equation}
    \varepsilon_{\pm}=\frac{M}{\left(1+\frac{e^{2}\gamma_{t}^{2}}{\mu^{2}}\right)}\left(\frac{\gamma_{s}}{\mu}\frac{e\gamma_{t}}{\mu}\pm\sqrt{\left(1+\frac{e^{2}\gamma_{t}^{2}}{\mu^{2}}\right)-\frac{\gamma_{s}^{2}}{\mu^{2}}}\right),
    \label{energia2}
\end{equation}
which is the expression associated with the energy spectrum of the generalized Coulomb potential obtained in \cite{barbosa2024}.

If we consider the special case where $\gamma_s=\gamma_t=\delta_r=\delta_t=0$ and without discarding the terms $\Delta_t^2$ and $\Delta_r^2$ in Equation~(\ref{eqveff}), we encounter the following energy spectrum
\begin{equation}
     \varepsilon_{\pm}^{2}=\pm \Bigg[ M^{2}+3 \Delta_r+\left(2+4N+\sqrt{1+\frac{4l_{\alpha}\left(l_{\alpha}+1\right)}{\beta^{2}}}\right)\sqrt{\Delta_r^2-\Delta_t^2}\Bigg],
\end{equation}
which is the energy spectrum for the Klein--Gordon equation with two nonminimal linear potentials in the double defect spacetime.

When we compute the energy spectrum under the assumptions $\gamma_s = \gamma_t = \delta_r = \delta_t = \Delta_t = 0$, substituting $\Delta_r = M \omega$ and retaining the $\Delta_r^2$ term in Equation~(\ref{eqveff}), we encounter the following:
\begin{equation}
    \varepsilon_{\pm}^{2}=\pm\left[M^{2}+2M\omega\left(2N'+\sqrt{\frac{l_{\alpha}\left(l_{\alpha}+1\right)}{\beta^{2}}+\frac{1}{4}}+\frac{1}{2}\right)\right],
\end{equation}
which is the energy spectrum for the Klein--Gordon oscillator in a spacetime with a cosmic string and a global monopole, where $N'=N+1$.
%By considering xx in equation (\ref{energia2})

\section{Discussion and Conclusions} \label{sec6}

In this paper we have discussed the spacetime with double topological defects, cosmic string plus global monopole, considering solutions of the Klein--Gordon equation. In this spacetime, the parameters \(\alpha\) and \(\beta\) represent the effects of the cosmic string and global monopole, respectively. It was observed that the quantum dynamics of a scalar particle is influenced by the double defects in a proper way. The spectrum of energy associated with the obtained solution of the Klein--Gordon equation undergoes a shift in its values, as can be seen from Equation~(\ref{energia}). The shift depends on the values of the defect parameters and the external potentials considered, demonstrating how the interaction of a particle with a potential plays an important role in defining quantum energy levels.  

In addition, we analyzed the scattering behavior of the quantum system and found that the phase shifts in the scattering process are influenced by the defect parameters and the presence of external potentials. The combined effect of the cosmic string and global monopole leads to modifications in the S-matrix, that encapsulates the phase shifts induced by the Coulomb and Cornell potentials, providing important information about the scattering process of the system.

We have analyzed in detail the behavior of the wave function and the radial probability density as a function of the radial coordinate. We tested several values for the potentials and analyzed the effects on the wave function. As a result, we have observed that the values for the interaction intensity change the values of the amplitude of the wave function and consequently of the probability density. 

Through the formalism used, the bound states can be obtained through the poles of the S-matrix. Thus, our results encompass both states with discrete spectrum and states that represent particles scattered by the potentials. Furthermore, due to the general characteristics of our system, some particular cases are included. We have treated the particular cases where $\Delta_r = \Delta_t = 0$, corresponding to the Cornell potential without linear term, which describes a system under the effect of Coulomb-type potentials. Another case of interest addressed corresponds to $\gamma_s=\gamma_t=\delta_r=\delta_t=0$, which is associated with the Klein--Gordon equation with two nonminimal linear potentials. In this way, our results are consistent with known limiting cases and the methods and findings presented here lay the groundwork for further studies on quantum behavior in nontrivial geometries, with possible extensions to other fields and interaction types.

\begin{acknowledgements}
LGB and JVZ acknowledge the financial support from CAPES (process numbers 88887.968290/2024-00 and 88887.655373/2021-00, respectively). LCNS would like to thank FAPESC for financial support under grant 735/2024.
\end{acknowledgements}

\bibliographystyle{spphys}
\bibliography{referencias_unificadas2}
\end{document}